\documentclass[a4paper,11pt]{article}
\pdfoutput=1
\usepackage{jheppub}
\usepackage[T1]{fontenc}
\usepackage{amsmath}
\usepackage{mathtools}
\newcommand{\SU}{\mathrm{SU}}
\newcommand{\U}{\mathrm{U}}

\newcommand{\dd}{{\rm{d}}}
\newcommand{\Tr}{{\rm Tr\,}}

\newcommand{\SW}{S_{\mbox{\tiny{W}}}}

\newcommand{\THagedorn}{T_{\mbox{\tiny{H}}}}
\newcommand{\TNG}{T_{\mbox{\tiny{NG}}}}
\newcommand{\mthreshold}{m_{\mbox{\tiny{th}}}}

\newcommand{\Up}{U_{\mbox{\tiny{p}}}}

\newcommand{\nconf}{n_{\mbox{\tiny{conf}}}}

\newcommand{\sigmaeff}{\sigma_{\mbox{\tiny{eff}}}}
\newcommand{\Lheat}{L_{\mbox{\tiny{h}}}}
\newcommand{\npar}{n_{\mbox{\tiny{par}}}}

\newcommand{\rmin}{r_{\mbox{\tiny{min}}}}
\newcommand{\redchisq}{\chi^2_{\tiny\mbox{red}}}
\newcommand{\eq}{\begin{equation}}
\newcommand{\en}{\end{equation}}
\newcommand{\eqar}{\begin{eqnarray}}
\newcommand{\enar}{\end{eqnarray}}

\title{\boldmath Hagedorn spectrum and thermodynamics of $\SU(2)$ and $\SU(3)$ Yang-Mills theories}

\author{Michele Caselle,}
\author{Alessandro Nada}
\author{and Marco Panero}

\affiliation{Department of Physics, University of Turin \& INFN, Turin \\
Via Pietro Giuria 1, I-10125 Turin, Italy}

\emailAdd{caselle@to.infn.it}
\emailAdd{anada@to.infn.it}
\emailAdd{marco.panero@unito.it}

\abstract{We present a high-precision lattice calculation of the equation of state in the confining phase of $\SU(2)$ Yang-Mills theory. We show that the results are described very well by a gas of massive, non-interacting glueballs, provided one assumes an exponentially growing Hagedorn spectrum. The latter can be derived within an effective bosonic closed-string model, leading to a parameter-free theoretical prediction, which is in perfect agreement with our lattice results. Furthermore, when applied to $\SU(3)$ Yang-Mills theory, this effective model accurately describes the lattice results reported by Bors{\'a}nyi et al. in JHEP 07 (2012) 056.}


\arxivnumber{1505.01106}

\begin{document}

\maketitle
\flushbottom

\section{Introduction}
\label{sec:introduction}

Working out the implications of Quantum Chromodynamics (QCD) for strongly interacting matter under extreme conditions of temperature and/or baryon density is one of the most important theoretical challenges in elementary particle physics~\cite{Fukushima:2011jc, Brambilla:2014jmp, Akiba:2015jwa, Muller:2015jva, Heinz:2015tua}. Such conditions, which are produced in high-energy nuclear collision experiments (for a recent review, see ref.~\cite{Andronic:2014zha}), have a dramatic \emph{qualitative} impact on the physics, as they directly affect the two most important phenomena determining the hadronic spectrum: confinement of colored degrees of freedom and dynamical breakdown of chiral symmetry. When the temperature is increased to values larger than approximately $160$~MeV, hadrons cease to exist, and QCD undergoes a crossover to a deconfined plasma phase, in which quarks and gluons interact with each other through screened long-range forces~\cite{Shuryak:1978ij}. When finite values of the quark chemical potential $\mu$ are allowed, additional phases (possibly characterized by phenomena like color superconductivity and superfluidity, color-flavor locking, et c.) may appear. 

Close to the deconfinement region, the quark-gluon plasma (QGP) produced in collider experiments appears to be (quite) strongly coupled, hence its theoretical description requires intrinsically non-perturbative tools. One possibility is provided by computations based on the gauge/string duality~\cite{Maldacena:1997re, Gubser:1998bc, Witten:1998qj}, which yields a description of the large-$N$~\cite{'tHooft:1973jz} and strong-coupling limits of a gauge theory, in terms of the classical gravity limit of a dual string model, defined in a higher-dimensional spacetime. The applications of holography to model the physics of hot QCD are discussed in detail in ref.~\cite{CasalderreySolana:2011us}. Another way to address the problem is via numerical Monte Carlo calculations in the lattice regularization of QCD~\cite{Wilson:1974sk}, which provides a gauge-invariant, non-perturbative definition of the theory. During the past decade, lattice QCD studies have provided conclusive quantitative answers to many questions relevant for the QGP~\cite{Philipsen:2012nu, Szabo:2014iqa, Ding:2015ona}: these include the nature of the deconfinement and chiral-symmetry-restoration crossover~\cite{Aoki:2006we}, the equation of state at vanishing chemical potential~\cite{Borsanyi:2013bia, Bazavov:2014pvz, Bhattacharya:2014ara}, the effect of electromagnetic fields on the QGP~\cite{Bali:2011qj, D'Elia:2012tr} and the determination of the freeze-out conditions relevant for heavy-ion collisions~\cite{Borsanyi:2011sw, Karsch:2012wm, Bazavov:2012jq, Borsanyi:2014ewa}. Although the lattice study of problems involving some form of dynamical evolution in real time is conceptually and practically much more challenging, recent works have addressed QGP phenomena related to transport and diffusion~\cite{Meyer:2007ic, Meyer:2007dy, CaronHuot:2009uh, Meyer:2010tt, Meyer:2011gj, Ding:2012sp, Aarts:2014nba}, to jet quenching and like-cone dynamics~\cite{Majumder:2012sh, Benzke:2012sz, Laine:2012ht, Ghiglieri:2013gia, Laine:2013lia, Laine:2013apa, Panero:2013pla, D'Onofrio:2014qxa, Brandt:2014uda}, and more.

Besides quantities of most direct phenomenological interest, the constant progress in algorithmic sophistication and in computational power makes it possible to address more fundamental issues, and to generalize the lattice studies of the QGP to QCD-like theories which, although not realized in nature, can provide helpful insight into the structure of non-Abelian gauge theories and can be compared, for instance, to analytical computations based on semiclassical approaches~\cite{Poppitz:2012nz}. Examples include finite-temperature lattice studies in pure-glue $\SU(3)$ Yang-Mills theory~\cite{Boyd:1996bx, Umeda:2008bd, Meyer:2009tq, Borsanyi:2012ve, Asakawa:2013laa, Giusti:2014ila, Francis:2015lha}, in its generalizations with a larger number of color charges $N$~\cite{Lucini:2002ku, Lucini:2003zr, Lucini:2005vg, Bursa:2005yv, Bringoltz:2005rr, Bringoltz:2005xx, Panero:2008mg, Panero:2009tv, Datta:2009jn, Datta:2010sq, Mykkanen:2012ri, Lucini:2012wq} and/or to three, rather than four, spacetime dimensions~\cite{Christensen:1991rx, Holland:2005nd, Holland:2007ar, Liddle:2008kk, Bialas:2008rk, Caselle:2011fy, Caselle:2011mn, Bialas:2012qz}, or theories based on a center-less exceptional gauge group~\cite{Pepe:2006er, Cossu:2007dk, Wellegehausen:2009rq, Bruno:2014rxa, Bonati:2015uga}. 

Following this line of research, in this article we report a high-precision study of the thermodynamics of $\SU(2)$ Yang-Mills theory (in four spacetime dimensions), focusing on the equilibrium properties in the confining phase. In particular, we show that the equation of state throughout the confining phase can be accurately modeled in terms of a gas of massive, non-interacting glueballs, provided that, in addition to the lightest glueball states (known from previous lattice calculations~\cite{Teper:1998kw}), one includes the contribution due to heavier glueballs, which can be described in terms of a Hagedorn stringy spectrum. The possibility that glueballs admit a description in terms of some sort of ``strings'' is quite natural, if one thinks of them in terms of ``rings of glue'', i.e. of closed color flux tubes. Historically, one of the first concrete realizations of this idea was put forward with the Isgur-Paton model~\cite{Isgur:1984bm}.

The present work, carried out according to the setup defined in section~\ref{sec:setup}, provides a direct generalization of the seminal lattice study presented in ref.~\cite{Meyer:2009tq} for $\SU(3)$ Yang-Mills theory, and later extended to $\SU(N)$ theories in three spacetime dimensions in ref.~\cite{Caselle:2011fy}. An important difference with respect to the three-color case in four spacetime dimensions, is that $\SU(2)$ Yang-Mills theory has a second-order deconfining phase transition; hence, if a Hagedorn temperature exists, it should coincide with the critical deconfinement temperature $T_c$. This constraint makes the heuristic glueball model in terms of closed bosonic strings much more predictive. As it will be shown in section~\ref{sec:results}, our lattice results for thermodynamic quantities in $\SU(2)$ Yang-Mills theory are in excellent agreement with the theoretical predictions of this phenomenological string model. Furthermore, in section~\ref{sec:results} we show that the same agreement also holds for the $\SU(3)$ theory: the bosonic string model allows one to derive a parameter-free prediction for the Hagedorn temperature, and including the contributions from the lightest glueball masses reported in ref.~\cite{Meyer:2004gx} one can obtain model predictions in excellent agreement with the results for thermodynamic quantities computed in ref.~\cite{Borsanyi:2012ve}. The presence of a Hagedorn spectrum~\cite{Hagedorn:1965st} in the low-energy sector of a theory is generally regarded as evidence for string-like dynamics (a very partial list of articles relevant for this subject includes refs.~\cite{Atick:1988si, Vafa:1995bm, Barbon:1998cr, Sundborg:1999ue, Aharony:2003sx, NoronhaHostler:2008ju, Haehl:2014yla, Belin:2014fna, Cohen:2015hwa}): the implications of the present work in this respect will be discussed in section~\ref{sec:conclusions}, together with some concluding remarks.

\section{Definitions and setup of the lattice calculation}
\label{sec:setup}

In this work, the regularization of $\SU(2)$ Yang-Mills theory in four-dimensional Euclidean spacetime is carried out in a finite hypercubic lattice $\Lambda$ of spacing $a$ and hypervolume $L_s^3 \times L_t = a^4 (N_s^3 \times N_t)$, in which periodic boundary conditions are imposed along all directions. As usual in thermal field theory, the shortest compactification size (which we take to be $L_t$) defines the temperature $T$ of the system via the relation $T=1/L_t$, while the sizes of the system in the three other, ``spatial'', directions are taken to be much larger ($9 \leq L_s/L_t \leq 12$) to avoid finite-volume effects.

The Euclidean action for the lattice theory is taken to be the standard Wilson action~\cite{Wilson:1974sk}
\begin{equation}
\label{wilson_action}
\SW = -\frac{2}{g^2} \sum_{x \in \Lambda} \sum_{0 \le \mu < \nu \le 3} \Tr U_{\mu\nu} (x)
\end{equation}
where $g$ denotes the (bare) lattice coupling and 
\begin{equation}
U_{\mu\nu} (x) = U_\mu (x) U_\nu \left(x+a\hat{\mu}\right) U_{\mu}^\dagger \left(x+a\hat{\nu}\right) U_{\nu}^\dagger (x)
\end{equation}
is the plaquette, with $U_\mu (x)$ an $\SU(2)$ group element (in the defining representation) which represents a parallel transporter (in color space) from the site $x$ to the site $x+a\hat{\mu}$. In the following, we also introduce the Wilson action parameter $\beta=4/g^2$.

At the quantum level, the dynamics of the lattice system is defined by the integral
\begin{equation}
Z = \int \prod_{x \in \Lambda} \prod_{\mu = 0}^{3} \dd U_\mu(x) \exp \left( -\SW \right)
\end{equation}
(where $\dd U_\mu$ denotes the $\SU(2)$ Haar measure), so that the expectation value of a generic, gauge-invariant quantity $\mathcal{A}$ is given by
\begin{equation}
\label{vev}
\langle \mathcal{A} \rangle = \frac{1}{Z} \int \prod_{x \in \Lambda} \prod_{\mu = 0}^{3} \dd U_\mu(x) \mathcal{A} \exp \left( -\SW \right).
\end{equation}
In our work, all expectation values of this form are estimated numerically, via Monte~Carlo integration: this is done by averaging over large sets of lattice configurations, produced by an algorithm based on a sequence of heat-bath~\cite{Creutz:1980zw, Kennedy:1985nu} and overrelaxation updates~\cite{Adler:1981sn, Brown:1987rra}. The statistical uncertainties due to the finiteness of the configuration sets are evaluated using the jackknife method~\cite{bootstrap_jackknife_book}.

We define the normalized, traced Polyakov loop through a point $x$ in the $t=0$ Euclidean time-slice $\Lambda_{t=0}$ of the lattice as
\begin{equation}
P(x) = \frac{1}{2} \Tr L_0 (x),
\end{equation}
where
\begin{equation}
L_0 (x) = \prod_{0 \le n < N_t} U_0 (x + n a \hat{0}).
\end{equation}
We computed the zero-temperature, two-point Polyakov loop correlation function
\begin{equation}
G(r) = \left\langle \frac{1}{3N_s^3} \sum_{x \in \Lambda_{t=0}} P(x) \sum_{1 \le i \le 3} P\left( x + r \hat{\imath} \right) \right\rangle
\end{equation}
at different values of $r$ and $\beta$ using the multilevel algorithm~\cite{Luscher:2001up} (see also refs.~\cite{Meyer:2003hy, Kratochvila:2003zj, Mykkanen:2012dv}) and extracted the interquark potential $V(r)$ from
\begin{equation}
\label{iqpotential}
V(r) = - \frac{1}{L_t} \ln G(r);
\end{equation}
these computations were carried out on lattices with $L_t =  32a$. This non-perturbative determination of $V(r)$ allows one to set the scale of the lattice simulations, extracting the value of the lattice spacing by fitting $V(r)$ to the following functional form:
\begin{equation}
\label{cornell}
a V = a \sigma r + a V_0 - \frac{\pi a}{12 r}.
\end{equation}
While this form is reminiscent of the old phenomenological Cornell potential~\cite{Eichten:1978tg} (up to the additive constant $aV_0$, which simply accounts for an overall renormalization), it is worth pointing out that eq.~(\ref{cornell}) can be derived in a more modern approach, assuming that color confinement is associated with the formation of a thin flux tube behaving like a quantum mechanical string, and carrying out an expansion around the low-energy (i.e., long-string) limit~\cite{Aharony:2013ipa}. In this approach, the coefficient of the $1/r$ term is unambiguously fixed by the massless nature of transverse string fluctuations~\cite{Luscher:1980fr}. The symmetries of the effective string action also fix the coefficients of further, subleading terms which could appear on the right-hand side of eq.~(\ref{cornell}), which are proportional to higher powers of $1/r$; however, we do not include them in eq.~(\ref{cornell}) because their contribution to the potential is negligible, within the precision of our numerical data.

The value of the lattice spacing in ``physical'' units can then be deduced from the fitted value of $\sigma a^2$, at each value of $\beta$. A complete determination of the relation between the lattice spacing $a$ and the Wilson action parameter $\beta$ for $\SU(2)$ pure gauge theory is discussed in section~\ref{sec:results}. For a detailed discussion of alternative scale-setting methods in lattice gauge theory, see ref.~\cite{Sommer:2014mea} and references therein.

A quantity of major phenomenological interest in finite-temperature field theory is the pressure $p$, which in the thermodynamic limit $V \to \infty$ equals the opposite of the free-energy density per unit volume $f$:
\begin{equation}
\label{pressuretdyn}
p = -\lim_{V \to \infty} f = \lim_{V \to \infty} \frac{T}{V} \ln{Z}.
\end{equation}
$p$ is related to the trace anomaly of the theory (denoted as $\Delta$) via
\begin{equation}
\frac{\Delta}{T^4} = T \frac{\partial}{\partial T} \left( \frac{p}{T^4} \right).
\end{equation}
Other two, closely related, quantities are the energy density per unit volume
\begin{equation}
\epsilon = \frac{T^2}{V} \left. {\frac{\partial \ln Z}{\partial T}}\right|_V = \Delta + 3p
\end{equation}
and the entropy density per unit volume
\begin{equation}
s = \frac{\epsilon}{T} + \frac{\ln Z}{V} = \frac{\Delta + 4p}{T}.
\end{equation}

Following the method introduced in ref.~\cite{Engels:1990vr}, it is straightforward to express these thermodynamic quantities in terms of plaquette expectation values: for example, the pressure (more precisely: its difference with respect to the value it takes at $T=0$) can be evaluated using
\begin{equation}
\label{pressure1}
p = \frac{6}{a^4} \int_{\beta_0}^{\beta} \dd \beta' \left( \langle \Up \rangle_T -  \langle \Up \rangle_0 \right),
\end{equation}
where $\langle \Up \rangle_T$ denotes the average plaquette at a generic temperature $T$. In practice, in the simulations one computes the integral appearing on the right-hand side of eq.~(\ref{pressure1}) numerically, typically using one of the methods discussed in ref.~\cite[appendix A]{Caselle:2007yc}, choosing a lower integration extremum $\beta_0$ deep enough in the confined phase, corresponding to a value of the temperature where the difference of the equilibrium thermodynamic quantities with respect to their zero-temperature values is negligible. 

Note that the integrand appearing in eq.~(\ref{pressure1}) is closely related to the trace anomaly,
\begin{equation}
\label{trace1}
\Delta = \frac{6}{a^4} \frac{\partial \beta}{\partial \ln a} \left( \langle \Up \rangle_0 -  \langle \Up \rangle_T \right),
\end{equation}
up to a factor which can be readily evaluated from the scale setting.

\section{Numerical results and comparison with a bosonic string model}
\label{sec:results}

In this section, we first present our results for the $\SU(2)$ Yang-Mills theory in subsection~\ref{subsec:su2results}, comparing them with the predictions of an effective model for the glueball spectrum, in which the physical states are interpreted as closed bosonic strings. Then, in subsection~\ref{subsec:su3results} we extend the analysis to the $\SU(3)$ results reported in ref.~\cite{Borsanyi:2012ve} (for the equation of state) and in ref.~\cite{Meyer:2004gx} (for the glueball masses).
 
\subsection{Results for $\SU(2)$ Yang-Mills theory}
\label{subsec:su2results}

We computed the static interquark potential for distances up to $r=16a$. The lower value of the distance range fitted to eq.~(\ref{cornell}) was chosen in such a way that $r \sqrt{\sigma} > 1$. The tree-level improved definition of the distance $r$ for the lattice potential introduced in ref.~\cite{Necco:2001xg} has been used.

Our results for the string tension in lattice units $\sigma a^2$ are reported in table~\ref{tab:stringtension}. For an accurate non-perturbative determination of the relation between $a$ and $\beta$, they have to be interpolated by a suitable functional form. 
One possibility is to fit the data for the logarithm of the string tension to a polynomial of degree $\npar - 1$ in $(\beta-\beta_0)$, with $\beta_0 = 2.4$.
Interpolating with
\begin{equation}
\label{betaform}
\ln (\sigma a^2) = \sum_{j=0}^{\npar-1} a_j (\beta-\beta_0)^j
\end{equation}
and choosing $\npar = 4$ the fit yields a $\redchisq$ of 0.01. We found the values of the parameters to be $a_0 = -2.68 $, $a_1 = -6.82 $, $a_2 = -1.90 $ and $a_3 = 9.96 $ respectively.
A fit with the functional form proposed in ref.~\cite{Allton:1996dn} yields a very small $\redchisq$ as well and similar results.

\begin{table}
\centering
\begin{tabular}{|c|c|c|c|c|}
\hline
$\beta$ & ${\rmin}/a$ & $\sigma a^2$ & $a V_0$ & $\redchisq$ \\
\hline \hline
$2.27 $ & $2.889$  & $0.157(8) $ & $ 0.626(14) $ & $ 0.6 $ \\
\hline
$2.30 $ & $2.889$  & $0.131(4) $ & $ 0.627(30) $ & $ 0.1 $ \\
\hline
$2.32 $ & $3.922$  & $0.115(6) $ & $ 0.627(32) $ & $ 2.3 $ \\
\hline
$2.35 $ & $3.922$  & $0.095(3) $ & $ 0.623(20)  $ & $ 0.2 $ \\
\hline
$2.37 $ & $3.922$  & $0.083(3) $ & $ 0.621(18)  $ & $ 1.0 $ \\
\hline
$2.40 $ & $4.942$  & $0.068(1) $ & $ 0.617(10)  $ & $ 1.4 $ \\
\hline
$2.42 $ & $4.942$ & $0.0593(4) $ & $ 0.613(5)  $ & $ 0.1 $ \\
\hline
$2.45 $ & $4.942$ & $0.0482(2) $ & $ 0.608(4)  $ & $ 0.4 $ \\
\hline
$2.47 $ & $4.942$ & $0.0420(4) $ & $ 0.604(5) $ & $ 0.3 $ \\
\hline
$2.50 $ & $5.954$ & $0.0341(2) $ & $ 0.599(2)  $ & $ 0.1 $ \\
\hline
$2.55 $ & $6.963$ & $0.0243(13) $ & $ 0.587(11) $ & $ 0.2 $ \\
\hline
$2.60 $ & $7.967$ & $0.0175(16) $ & $ 0.575(16) $ & $ 0.3 $ \\
\hline
\end{tabular}
\caption{\label{tab:stringtension} Results for the string tension (in units of the inverse squared lattice spacing, in the third column) at different values of the Wilson action parameter $\beta$ (first column), calculated by fitting the potential $V$ as a function of the tree-level improved interquark distance $r$ (see ref.~\cite{Necco:2001xg} for details) to eq.~(\protect\ref{cornell}). $V$ was extracted from Polyakov loop correlators on lattices of temporal extent $L_t=32a$. The minimal distances (reported in the second column, in units of the lattice spacing) used in the fits were fixed by the $r \sqrt{\sigma} > 1$ condition. The table also shows the results for the ``perimeter-like'' term $a V_0$ (fourth column) and the $\redchisq$ values (fifth column).}
\end{table}

\begin{figure}
\begin{center}
\includegraphics*[width=\textwidth]{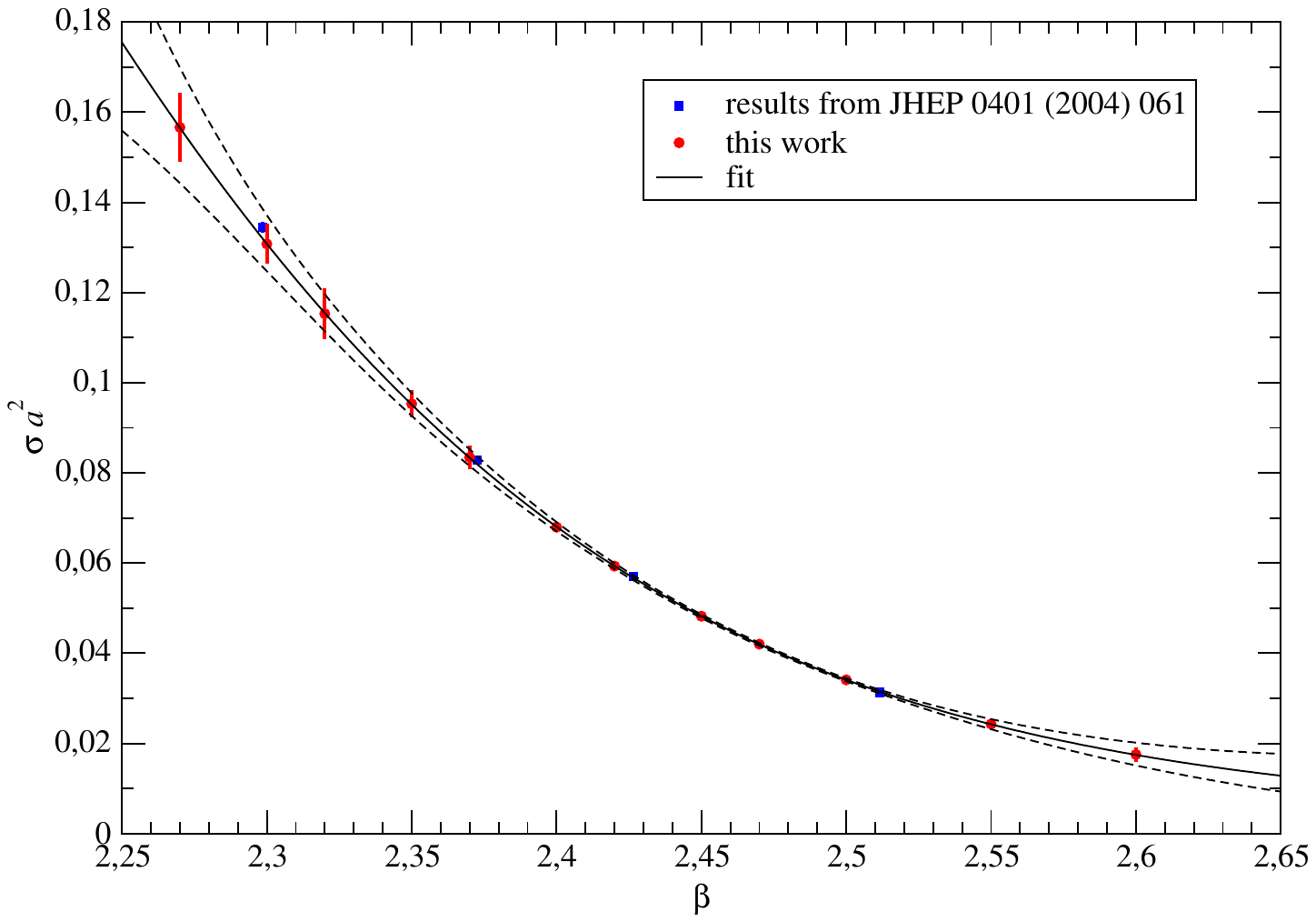}
\caption{\label{fig:scalesetting} The values of the string tension in lattice units obtained from our lattice simulations (red circles) are showed along with those reported in ref.~\cite{Lucini:2003zr} (blue squares). The solid black curve and the dashed black lines show the interpolation to the functional form in eq.~(\protect\ref{betaform}), with the associated uncertainties.}
\end{center}
\end{figure}

This scale setting allows one to get an accurate determination of the temperature for a given value of the parameter $\beta$.

\begin{figure}
\begin{center}
\includegraphics*[width=\textwidth]{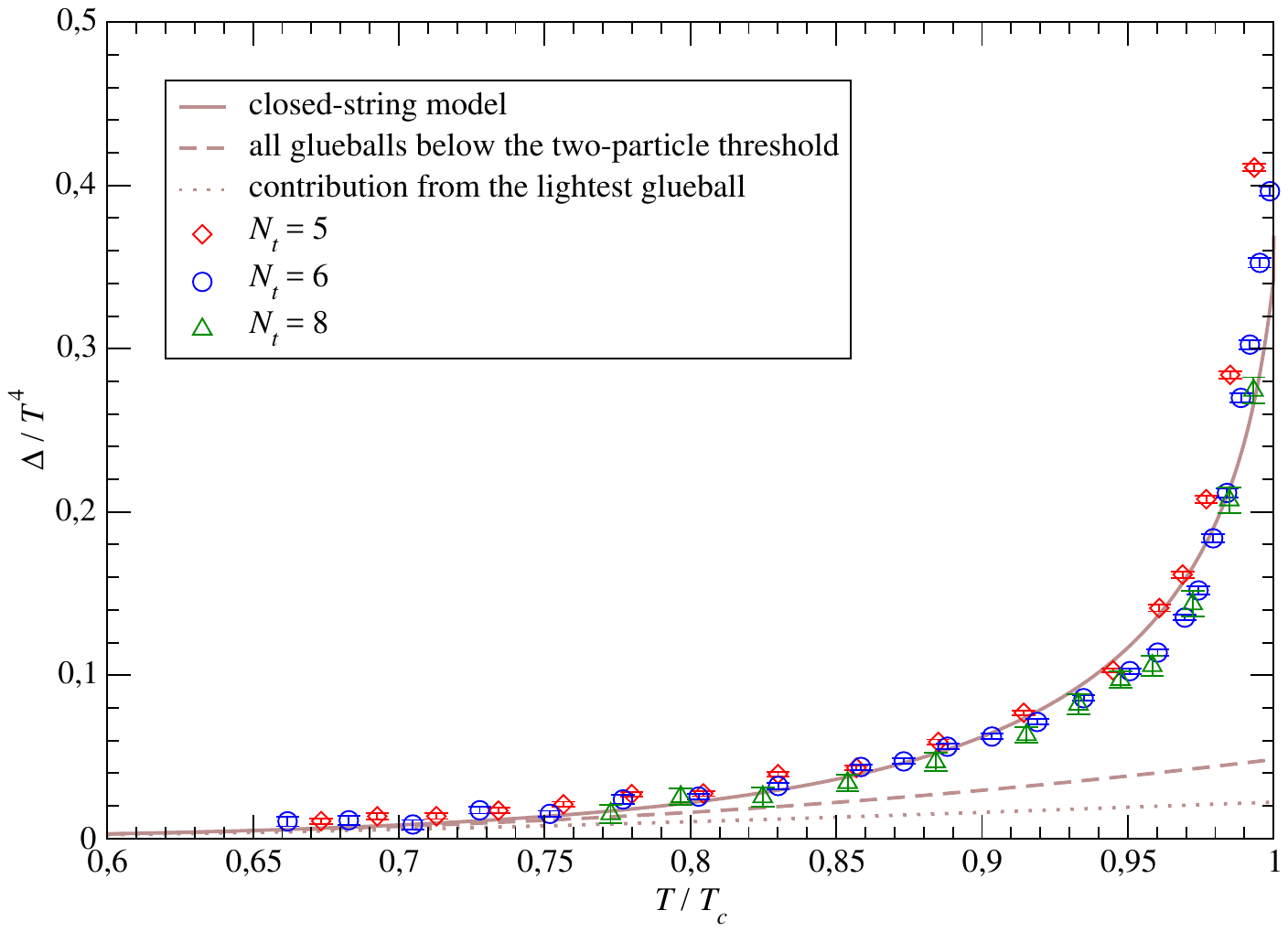}
\caption{\label{fig:su2_trace} Comparison between our lattice results for the trace anomaly in $\SU(2)$ Yang-Mills theory from simulations with $N_t=5$, $6$ and $8$ (brown symbols) and the behavior expected for a gas of free, massive glueballs. The dotted line corresponds to the contribution of the lightest state only, with quantum numbers $J^{PC}=0^{++}$. The dashed line includes all the low-lying glueballs, with masses below the threshold associated with the appearance of two $0^{++}$ glueballs  (taken from \cite{Teper:1998kw}) while the solid line corresponds to the case in which we included also the contribution from high-lying states, described by a bosonic string model. The trace anomaly $\Delta$ is displayed in units of the fourth power of the temperature, and is plotted as a function of $T/T_c$.}
\end{center}
\end{figure}

The main part of our study of $\SU(2)$ Yang-Mills theory consists in a high-precision non-perturbative determination of the finite-temperature equation of state in the confining phase. Table~\ref{tab:thermsetup} shows the setup of Monte~Carlo simulations to compute the thermodynamic observables defined in section~\ref{sec:setup}. The results obtained from lattices with temporal size (in lattice-spacing units) 
$N_t = 5$, $6$ and $8$ are showed in figure~\ref{fig:su2_trace}. 
Note that on the horizontal axis the temperature is displayed in units of the deconfinement temperature $T/T_c$, that we derived using the value 
for $T_c/\sqrt{\sigma} = 0.7091(36)$ reported in ref.~\cite{Lucini:2003zr}.

\begin{table}
\centering
\begin{tabular}{|c|c|c|c|c|}
\hline
$N_s^4$ at $T=0$ & $N_s^3 \times N_t$ at $T\neq 0$ & $n_\beta$ & $\beta$-range & $\nconf$\\
\hline \hline
$32^4$ & $60^3 \times 5$ & $17$ & $[2.25,2.3725]$ & $1.5 \times 10^5$ \\
\hline
$40^4$ & $72^3 \times 6$ & $25$ & $[2.3059,2.431]$ & $1.5 \times 10^5$ \\
\hline
$40^4$ & $72^3 \times 8$ & $12$ & $[2.439,2.5124]$ & $10^5$ \\
\hline
\end{tabular}
\caption[Lattice setup for $\SU(2)$ simulations]{\label{tab:thermsetup} Setup of our simulations. The first two columns show the lattice sizes (in units of the lattice spacing $a$) for the $T=0$ and finite-temperature simulations, respectively. In the third column, $n_\beta$ denotes the number of $\beta$-values simulated within the $\beta$-range indicated in the fourth column. Finally, in the fifth column we report the cardinality $\nconf$ of the configuration set for the $T=0$ and finite-$T$ simulations.}
\end{table}

Because of confinement and dynamical generation of a finite mass gap, the only physical states in the confined phase of $\SU(N)$ Yang-Mills theory are massive glueballs. Under the assumption that such states are weakly interacting with each other (which is supported by theoretical arguments~\cite{Mathieu:2008me} and appears to be compatible with experimental results in real-world QCD~\cite{Crede:2008vw}), it is reasonable to think that the thermodynamic observables of these theories can be modelled in terms of a gas of free, relativistic bosons.

The pressure of a gas of non-interacting, relativistic bosons of mass $m$ is given by
\begin{equation}
\label{B1}
p = \frac{m^2T^2}{2\pi^2} \sum_{n=1}^\infty \frac{K_2 \left( nm/T \right)}{n^2},
\end{equation}
where $K_\nu(z)$ denotes a modified Bessel function of the second kind of index $\nu$. Correspondingly, the trace anomaly reads
\begin{equation}
\Delta = \frac{m^3T}{2 \pi^2} \sum_{n=1}^\infty \frac{K_1 \left( nm/T \right)}{n}.
\label{B2}
\end{equation}
Using the fact that in the confining phase $T < T_c$, and that, in turn, all glueball masses are significantly larger than $T_c$, the expressions above can be recast into a simpler form, using known asymptotic expressions for the Bessel functions $K_\nu(z)$, which are valid for large real values of $z$:
\begin{equation}
p \simeq T \left( \frac{T m}{2 \pi} \right)^{3/2} \sum_{n=1}^\infty \frac{\exp \left(-nm/T \right)}{n^{5/2}} \left( 1 + \frac{15T}{8nm} \right)
\label{exp1}
\end{equation}
and
\begin{equation}
\Delta \simeq m \left( \frac{T m}{2 \pi} \right)^{3/2} \sum_{n=1}^\infty \frac{\exp \left(-nm/T \right)}{n^{3/2}} \left( 1 + \frac{3T}{8nm} \right).
\label{exp2}
\end{equation}
In particular, the precision of our numerical results is sufficient to investigate the first subleading terms in the asymptotic expansions in eq.~(\ref{exp1}) and in eq.~(\ref{exp2}), so we truncate the expansions accordingly.

Fig.~\ref{fig:su2_trace} displays our lattice results for $\Delta/T^4$ as a function of $T/T_c$ in $\SU(2)$ Yang-Mills theory, together with their comparison with the prediction from the glueball-gas model described by eq.~(\ref{glueballgas}), for which we used the results reported in ref.~\cite{Teper:1998kw} for the lightest spectrum states and the value of $T_c$ taken from ref.~\cite{Lucini:2003zr}.

In fig.~\ref{fig:su2_trace}, the dotted line shows the contribution to $\Delta/T^4$ from a glueball gas including only the lightest state, with quantum numbers $J^{PC}=0^{++}$, whereas the dashed line is obtained including all states with masses below the threshold associated with the appearance of two $0^{++}$ glueballs. The plot reveals that there is a large mismatch between the contribution of these light glueballs and our lattice results. 

Note that, due to the exponential dependence on the mass in eq.~(\ref{exp1}) and in eq.~(\ref{exp2}), this mismatch can only be accounted for, if heavier glueballs are described by a Hagedorn-like (i.e. exponentially increasing) spectrum. In a confining gauge theory, a Hagedorn-like spectrum does arise, when hadrons are modelled in terms of thin color flux tubes, whereby mesonic states (made of a bound quark-antiquark pair) can be described by an \emph{open} bosonic string~\cite{Luscher:1980fr}, while glueballs are described by a \emph{closed} bosonic string~\cite{Isgur:1984bm, Johnson:2000qz}.

To be more quantitative, the closed bosonic string model leads to a spectral density
\begin{equation}
 \hat{\rho} (m) = \frac{1}{m} \left( \frac{2\pi \THagedorn}{3m}\right)^{3} \exp \left( m/\THagedorn \right)
\end{equation}
(for a detailed derivation see ref.~\cite[appendix]{Caselle:2011fy}),
in which the only free parameter is the Hagedorn temperature $\THagedorn$. The latter, actually, can also be predicted by the string model, if the action governing the dynamics of the bosonic string is known. The simplest \emph{Ansatz} for the effective string action is the Nambu-Got{\={o}} action~\cite{Nambu:1974zg, Goto:1971ce}: although it is non-renormalizable and affected by problems at a fundamental level (including, for example, an anomaly), recent advances in our understanding of the effective string description for Yang-Mills theories indicate that it provides a very good approximation of the actual effective string action, up to very small corrections that appear only at a large order in the expansion around the long-string limit~\cite{Aharony:2013ipa}. Assuming that the glueball spectrum is described by a  closed Nambu-Got{\={o} string model, one obtains
\begin{equation}
\label{Hagedorntemp}
\THagedorn = \TNG =\sqrt{\frac{3 \sigma}{2\pi}} \simeq 0.691 \sqrt{\sigma}.
\end{equation}
If the deconfinement transition is of second order, as is the case in $\SU(2)$ Yang-Mills theory, then $\THagedorn$ coincides with the deconfinement temperature.\footnote{Note that the $\THagedorn$ value in eq.~(\ref{Hagedorntemp}) is slightly lower than the critical deconfinement temperature determined numerically in ref.~\cite{Lucini:2003zr} for this model: $T_c/\sqrt{\sigma} = 0.7091(36)$; albeit small (less than $3\%$), the difference is statistically significant. If taken at face value, this discrepancy would imply the surprising (and almost paradoxical) conclusion that the partition function of the $\SU(2)$ theory ought to diverge \emph{before} reaching the deconfinement transition---which is clearly unsupported by our simulations. Barring the possibility of additional, unknown uncertainties on the lattice results, the most likely solution of the conundrum is that the effective string model fails to capture the dynamics at temperatures very close to $T_c$. In fact, as it will be discussed in more detail in section~\ref{sec:conclusions}, it is known that, in the presence of a second-order deconfinement transition, the behavior of long-range correlation functions at criticality is consistent with the indices predicted by the Svetitsky-Yaffe conjecture~\cite{Svetitsky:1982gs}, rather than with the mean-field behavior that one would expect from a bosonic string. An alternative possibility is that, besides the Nambu-Got{\={o}} term, the actual effective string action describing the confined phase of the theory includes additional contributions, which do not spoil Lorentz invariance~\cite{Aharony:2011gb}. For example, using high-precision lattice calculations in the dual formulation of the model~\cite{Panero:2004zq, Panero:2005iu}, it was recently shown that the effective string action describing compact $\U(1)$ lattice gauge theory in three dimensions includes a non-negligible contribution from an extrinsic curvature term~\cite{Caselle:2014eka}. However, as discussed in detail in ref.~\cite{Caselle:2014eka}, the numerical relevance of this particular contribution is closely related to the non-trivial (and highly non-generic) scaling of the ratio between the square root of the string tension and the mass gap in compact $\U(1)$ lattice gauge theory in three dimensions, and we do not expect a similar effect to be relevant for $\SU(2)$ Yang-Mills theory. As a matter of fact, lattice calculations confirm that the Nambu-Got{\={o}} string provides a very accurate description of the low-energy dynamics of non-Abelian gauge theories, both in three and in four spacetime dimensions~\cite{Teper:2009uf, Brandt:2013eua}.} On the other hand, if the deconfinement transition is discontinuous, as in $\SU(N \ge 3)$ theories, then the Hagedorn temperature is expected to be larger than $T_c$.

The free glueball gas prediction for the trace anomaly can then be written as a sum of two contributions:
\begin{equation}
\label{glueballgas}
\Delta(T) = \sum_{m_i < \mthreshold} (2J+1) \Delta (m_i,T) + n_C \int_{\mthreshold}^{\infty} \dd m^\prime \hat{\rho}(m^\prime) \Delta(m^\prime,T).
\end{equation}
On the right-hand side, the first addend is the contribution from known glueballs of masses up to a threshold scale, which we conventionally assume to be $\mthreshold = 2 m_{0^{++}}$, while the second accounts for contributions from heavier states, described according to the spectral density of the bosonic string model. The multiplicity factor $n_C$ counts the number of charge-conjugation eigenvalues; for the $\SU(2)$ Yang-Mills theory, the pseudo-reality of the gauge group implies that only $C=1$ glueballs exist, hence $n_C$ is fixed to $1$.

As fig.~\ref{fig:su2_trace} shows, our data are in remarkable agreement with the prediction of the closed bosonic string model. We remark that this comparison has \emph{no free parameters} except for the choice of $\mthreshold = 2 m_{0^{++}}$, but we checked that our results are robust against small changes of this threshold. As such, this is a very stringent test of both the effective string model and of the assumption that the confining regime of $\SU(2)$ Yang-Mills theory is described well by a gas of massive, non-interacting glueballs.

The corresponding results for the pressure are shown in fig.~\ref{fig:su2_pressure}.

\begin{figure}
\begin{center}
\includegraphics*[width=\textwidth]{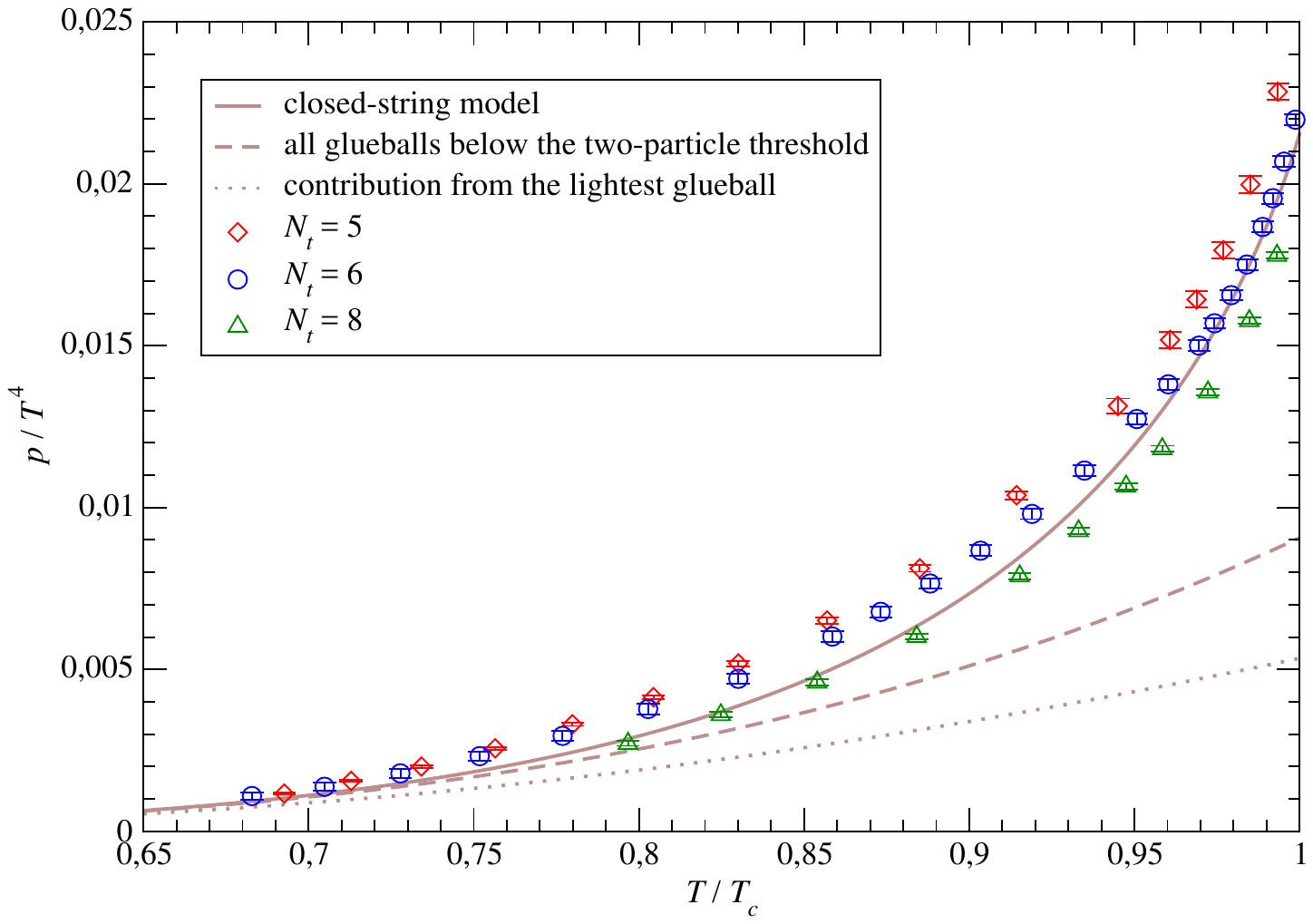}
\caption{\label{fig:su2_pressure} Same as in fig.~\protect\ref{fig:su2_trace}, but for the pressure $p$, in units of the fourth power of the temperature.}
\end{center}
\end{figure}

\subsection{Comparison with $\SU(3)$ Yang-Mills theory}
\label{subsec:su3results}

It is instructive to compare our findings with those obtained in Yang-Mills theory with $\SU(3)$ gauge group. To this end, we use the very precise data for $\Delta/T^4$ (extrapolated to the continuum limit) reported in ref.~\cite{Borsanyi:2012ve} and the glueball masses obtained in ref.~\cite{Meyer:2004gx}.

As discussed above, in contrast to the two-color case, the $\SU(3)$ theory features a first-order deconfinement transition, hence the Hagedorn temperature is expected to be different from the deconfinement temperature. We assume that the Hagedorn temperature coincides with the temperature at which the ``effective'', temperature-dependent string tension predicted by the Nambu-Got\={o} string model vanishes, namely $\THagedorn=\sqrt{3 \sigma/(2\pi)}$. An additional difference with respect to the $\SU(2)$ case, is that Yang-Mills theories with $N \ge 3$ color charges admit $C=-1$ glueballs (whereas for $N=2$ only $C=1$ states are allowed, since all irreducible representations of the gauge group are real or pseudo-real). In our analysis for the $\SU(3)$ theory, we include all $C=\pm 1$ states of masses below $\mthreshold$ and count the bosonic string contribution in eq.~(\ref{glueballgas}) twice, i.e. in eq.~(\ref{glueballgas}) $n_C$ is set to $2$.

It should be noted that, for the lightest states in the $\SU(3)$ spectrum, there is no mass degeneracy between glueballs of opposite charge-conjugation eigenvalues: the states in the $C=1$ sector are lighter than those with $C=-1$ and with the same $J$ and $P$ quantum numbers~\cite{Morningstar:1999rf, Meyer:2004gx}. While it may be that some hierarchy of this sort persists for heavier states, one notes that the spectra of the lightest $C=1$ and $C=-1$ glueballs are not just trivially shifted with respect to each other. On the other hand, the bosonic string model, \emph{per se}, does not suggest the existence of a finite gap between glueballs of different $C$ (at least for the part of the spectrum that it is expected to describe, i.e. heavier states). Hence, for the sake of simplicity, and to avoid introducing an arbitrary number of additional free parameters in our effective model, we assume that the states of opposite $C$ above the two-particle threshold are mass-degenerate---and interpret the gap between the lightest $C=1$ and $C=-1$ particles as an accidental feature of the lowest end of the spectrum (which certainly is not expected to be described by a model predicting a continuous density of states).\footnote{The idea that heavier glueball states of different $C$ may exhibit mass degeneracies, which are absent among the lightest states, should not be completely surprising. In QCD, due to the dynamical breaking of chiral symmetry at low energies and its approximate restoration at higher energies, excited hadrons of different parity $P$ tend to become mass-degenerate, while the effects of dynamical chiral-symmetry breaking are dramatic on the lightest states, and lead to large gaps between particles with $P=1$ and $P=-1$~\cite{Glozman:1999tk, Glozman:2007ek}.}

The results for both $\SU(2)$ and $\SU(3)$ Yang-Mills theories are plotted in fig.~\ref{fig:su2su3} as a function of $T/\THagedorn$: the numerical results are in excellent agreement with the predictions from the effective string model. Note that, as already observed in $2+1$ dimensions~\cite{Caselle:2011fy}, the presence of $C=-1$ states for $N>2$ plays a crucial r\^ole. We also remark that, using the string prediction $\THagedorn=\sqrt{3 \sigma/(2\pi)}$ for the $\SU(3)$ theory, the curves predicted by the string model do not depend on any free parameters: this is the case both for $N=2$ and $N=3$ colors.

\begin{figure}
\begin{center}
\includegraphics*[width=\textwidth]{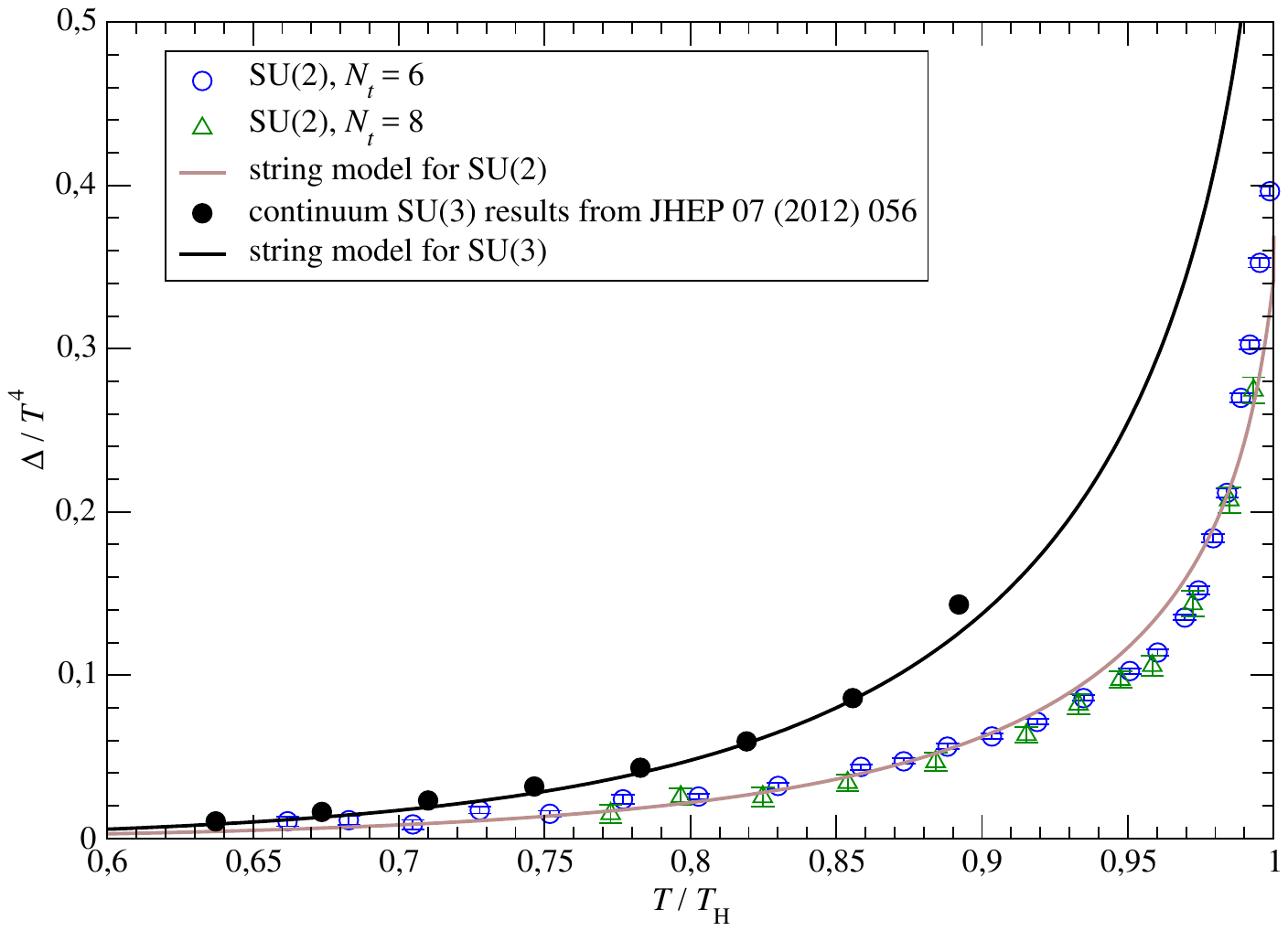}
\caption{\label{fig:su2su3} Comparison between the predictions of a massive-glueball gas, supplemented by the contribution from states modelled by a closed Nambu-Got\={o} string model, like in eq.~(\protect\ref{glueballgas}), and continuum-extrapolated data obtained in ref.~\cite{Borsanyi:2012ve} for $\SU(3)$ Yang-Mills theory and our data for the $\SU(2)$ theory (from simulations at finite lattice spacings corresponding to $N_t=6$ and $8$), as a function of $T/\THagedorn$.}
\end{center}
\end{figure}

\section{Discussion and concluding remarks}
\label{sec:conclusions}

In this work we presented some novel, high-precision numerical lattice results for the equation of state in $\SU(2)$ Yang-Mills theory. In particular, we studied the equilibrium thermodynamics in the confining phase, i.e. at temperatures below the critical temperature at which the second-order thermal deconfinement transition takes place. We showed that it can be described very well by a gas of free massive glueballs, provided one includes the contribution due to higher states in the spectrum, which we modelled using a bosonic closed-string model. We also showed that similarly good agreement holds for the results for the $\SU(3)$ theory obtained in ref.~\cite{Borsanyi:2012ve}, as already pointed out therein (see, in particular, ref.~\cite[fig.~4]{Borsanyi:2012ve}) and in ref.~\cite{Meyer:2009tq}. Note, however, that the analysis carried out in those works was slightly different: in particular, in ref.~\cite{Meyer:2009tq} the Hagedorn temperature was fitted from the numerical results and turned out to be about $4\%$ smaller than the Nambu-Got\={o} string estimate.\footnote{This mismatch with respect to our results might be due to a missing $n_C=2$ factor (which accounts for the existence of both $C=1$ and $C=-1$ glueballs) multiplying the closed bosonic string density of states: we verified numerically that setting $n_C=1$ in eq.~(\ref{glueballgas}) would lead to the same shift in $\THagedorn$.} A closely related analysis was presented in ref.~\cite{Buisseret:2011fq}, in which a generalization to other gauge groups was also discussed. Another phenomenological model to describe the Yang-Mills equation of state in terms of a glueball gas was discussed in ref.~\cite[section 7]{Arriola:2014bfa}. Another interesting lattice study of the Hagedorn temperature in $\SU(N)$ Yang-Mills theories at large $N$ was presented in ref.~\cite{Bringoltz:2005xx}, where the value of $\THagedorn$ was calculated by determining the temperature-dependent ``effective string tension'' $\sigmaeff$ associated with the two-point spatial correlation function of flux tubes. Such effective string tension is a decreasing function of $T$, and $\THagedorn$ can be identified as the temperature at which $\sigmaeff$ vanishes. Since $\SU(N \ge 3)$ Yang-Mills theories undergo a first-order thermal deconfining transition at a critical temperature $T_c < \THagedorn$, one cannot directly reach the point at which the spatial correlation length $1/\sqrt{\sigmaeff}$ diverges, but the key observation of ref.~\cite{Bringoltz:2005xx} is that one can, nevertheless, extract $\sigmaeff$ values even for $T > T_c$, as long as one remains in a \emph{metastable} confining phase. The existence of the latter follows from the strongly first-order nature of the deconfinement transition at large $N$~\cite{Lucini:2003zr, Lucini:2005vg}, it allows one to compute $\sigmaeff$ at temperatures sufficiently close to $\THagedorn$, and finally to carry out an accurate determination of $\THagedorn$ from a short-ranged extrapolation in $T$: the results indicate that $\THagedorn/T_c \simeq 1.1$ at large $N$, with some variation depending on the type of fit used for the extrapolation. Note that in our numerical study of $\SU(2)$ Yang-Mills theory, this procedure is not necessary, as the correlation length diverges at $T_c$. For the $\SU(3)$ theory, on the other hand, the deconfining transition is a discontinuous one and $\THagedorn > T_c$, but the latent heat $\Lheat$ associated with the deconfinement transition is smaller than at large $N$, and probing the metastability region may be more challenging.

Our findings add new, quantitative support to the idea that the low-energy dynamics of confining gauge theories admits an accurate description in terms of a bosonic string model. Loosely speaking, in this picture glueballs can be interpreted as ``rings of glue''~\cite{Isgur:1984bm}. The simplest assumption about the effective action describing the dynamics of these fluctuating loops of color flux is that it coincides with the Nambu-Got\={o} action. It is worth remarking that this assumption is not arbitrary: the effective theory describing the low-energy dynamics of confining gauge theories must be consistent with a non-linear realization of Lorentz-Poincar\'e symmetry~\cite{Aharony:2009gg, Aharony:2010cx, Aharony:2011gb, Aharony:2013ipa, Billo:2012da, Gliozzi:2011hj, Gliozzi:2012cx, Meyer:2006qx, Dubovsky:2012sh, Dubovsky:2013gi, Dubovsky:2014fma, Ambjorn:2014rwa, Caselle:2014eka}, a requirement imposing tight constraints not only on the functional form of the terms appearing in the effective action, but also on their \emph{coefficients}. In particular, the leading terms (in an expansion around the long-string limit) turn out to agree precisely with those of the Nambu-Got\={o} action. This is confirmed numerically by a vast body of lattice calculations, reviewed in refs.~\cite{Kuti:2005xg, Billo:2012da, Teper:2009uf, Lucini:2012gg, Panero:2012qx, Lucini:2013qja, Brandt:2013eua}.

A word of caution, however, is in order: the continuous nature of the thermal deconfining transition in $\SU(2)$ Yang-Mills theory implies that, at $T=T_c$, the dynamics of the system (including, for example, the way the spatial correlation length of Polyakov loops diverges) is characterized by the critical indices of the tridimensional Ising model~\cite{Svetitsky:1982gs}. By contrast, the Nambu-Got\={o} string model predicts mean-field exponents. This indicates that, strictly speaking, the effective string model cannot completely capture the dynamics of the system, all the way up to $T_c$. Nevertheless, our numerical results indicate that the agreement is very good over a broad range, reaching temperatures close to the deconfinement one. We note that analogous conclusions could be drawn from the study of $\SU(3)$ Yang-Mills theory in three dimensions reported in ref.~\cite{Bialas:2012qz}: this theory has a second-order deconfining transition, and the analysis carried out in ref.~\cite{Bialas:2012qz} showed that the Nambu-Got\={o} string model successfully predicts the value of the deconfinement temperature, even though the critical behavior very close to $T_c$ is consistent with the universality class of the bidimensional three-state Potts model (i.e. agrees with the Svetitsky-Yaffe conjecture) rather than with the mean-field exponents predicted by the bosonic string. Similar findings were also obtained from the study of large-$N$ Yang-Mills theories in the aforementioned ref.~\cite{Bringoltz:2005xx}, where it was found that, close to $\THagedorn$, the scaling of the effective spatial string tension extracted from Polyakov loops at large $N$ is better described by the critical exponent of the tridimensional XY model, than by the one predicted by a bosonic string. 

A natural implication of the Hagedorn scenario for the deconfinement transition is that glueballs, due to their divergent contribution to the equation of state, cannot exist at temperatures (much) above the deconfinement temperature~\cite{Caselle:2013qpa}. This non-trivial observation can be contrasted with the claims reported in refs.~\cite{Ishii:2002ww, Meng:2009hh}. 

While the idea that a thermal deconfining transition takes place in the presence of an exponentially growing, Hagedorn-like, hadronic spectrum dates back to forty years ago~\cite{Cabibbo:1975ig}, it should be pointed out that only recently has it become possible to test this conjecture \emph{quantitatively} from first principles, via high-precision Monte~Carlo computations on the lattice. In this respect, our present work can be considered as a generalization of recent studies, that addressed this problem either in $D=2+1$ (rather than $3+1$) spacetime dimensions~\cite{Caselle:2011fy}, or in $3+1$ spacetime dimensions, but for $N=3$ color charges~\cite{Meyer:2009tq, Borsanyi:2012ve}, with similar results. As we discussed, the theory with $N=2$ color charges addressed here provides an important cross-check of the glueball-gas model, given that the spectrum of $\SU(2)$ Yang-Mills theory does not contain $C=-1$ states.

The present work could be generalized along different directions: for example, one could extend the present analysis to $\SU(N>3)$ Yang-Mills theories, and/or take the effect of glueball interactions into account, possibly in terms of excluded-volume corrections~\cite{Kapusta:1982qd, Yen:1997rv}, as recently discussed in refs.~\cite{Albright:2014gva, Alho:2015zua}.

\acknowledgments
The simulations were performed on INFN Pisa GRID Data Center machines. We thank Keijo~Kajantie, Gwendolyn~Lacroix, Enrique~Ruiz~Arriola, and Giorgio~Torrieri for comments and discussions, and Etsuko~Itou for calling our attention to a misprint in the discussion of eq.~(\ref{betaform}), which has been corrected in the present version of the manuscript.

\bibliography{paper}

\end{document}